\documentclass[sigconf]{acmart}

\AtBeginDocument{%
  \providecommand\BibTeX{{%
    \normalfont B\kern-0.5em{\scshape i\kern-0.25em b}\kern-0.8em\TeX}}}

\copyrightyear{2026}
\acmYear{2026}
\setcopyright{cc}
\setcctype{by-nc-nd}
\acmConference[CHI '26]{Proceedings of the 2026 CHI Conference on Human Factors in Computing Systems}{April 13--17, 2026}{Barcelona, Spain}
\acmBooktitle{Proceedings of the 2026 CHI Conference on Human Factors in Computing Systems (CHI '26), April 13--17, 2026, Barcelona, Spain}
\acmPrice{}
\acmDOI{10.1145/3772318.3791632}
\acmISBN{979-8-4007-2278-3/2026/04}

\usepackage{longtable}
\usepackage{makecell}

\begin{document}

\title[What is \textit{Safety}? Corporate Discourse, Power, and the Politics of Generative AI Safety]{What is \textit{Safety}? Corporate Discourse, Power, and the Politics of Generative AI Safety}



\author{Ankolika De}
\email{apd5873@psu.edu}
\affiliation{%
  \institution{College of Information Sciences and Technology, The Pennsylvania State University}
  \country{USA}
}
\authornote{This work was conducted while the author was a visiting PhD Scholar at the Max Planck Institute for Security and Privacy.}

\author{Gabriel Lima}
\email{gabriel.lima@mpi-sp.org}
\affiliation{%
  \institution{Max Planck Institute for Security and Privacy}
  \country{Germany}
}

\author{Yixin Zou}
\email{yixin.zou@mpi-sp.org}
\affiliation{%
  \institution{Max Planck Institute for Security and Privacy}
  \country{Germany}
}
\renewcommand{\shortauthors}{De et al.}


\begin{abstract}
This work examines how leading generative artificial intelligence companies construct and communicate the concept of "safety" through public-facing documents. Drawing on critical discourse analysis, we analyze a corpus of \textit{corporate safety-related statements} to explicate how authority, responsibility, and legitimacy are discursively established. These discursive strategies consolidate legitimacy for corporate actors, normalize safety as an experimental and anticipatory practice, and push a perceived participatory agenda toward safe technologies. We argue that uncritical uptake of these discourses risks reproducing corporate priorities and constraining alternative approaches to governance and design. The contribution of this work is twofold: first, to situate safety as a sociotechnical discourse that warrants critical examination; second, to caution human-computer interaction scholars against legitimizing corporate framings, instead foregrounding accountability, equity, and justice. By interrogating safety discourses as artifacts of power, this paper advances a critical agenda for human-computer interaction scholarship on artificial intelligence.
\end{abstract}




\maketitle

\section{Introduction}

\begin{quote}
   \textit{ AI is one of the most important things humanity is working on. It is more profound than, I dunno, electricity or fire.}
    
    --- Sundar Pichai, CEO, Google \cite{clifford2018google}
\end{quote}

Artificial Intelligence (AI) technologies are indeed expanding rapidly in capability, availability, and integration across domains such as healthcare, education, finance, and creative industries, yet their impacts and trajectories remain far more complex and uncertain than such sweeping comparisons suggest. Instead, such analogies exemplify corporate discursive tendencies to situate AI as a transformative and universal force, one whose trajectory appears both natural and unavoidable.

Our work focuses on companies that have developed and deployed general-purpose chatbot-based Generative AI (GenAI), which has been widely adopted and appropriated by various actors, as it allows for \textit{conversations on a wide range of topics} \cite{trofymenko2021taxonomy}. These applications and their rapid, ongoing deployment have fueled debates about \emph{safety}, especially as real-world incidents underscore both the potential benefits and risks of AI deployment. Recent cases---including misinformation amplification, algorithmic bias, and psychological harms to individuals~\cite{10.1145/3689372, 10.1145/3706598.3713429, 10.1145/3715275.3732039, feehly2025truth}---underscore that safety is not only a technical concern but also a social and political one. 
Safety is a contested, context-dependent concept shaped by those with the power to define it \cite{ellcessor2022maps}, encompassing technical reliability, risk mitigation, but also social, ethical, and regulatory dimensions that determine who benefits from the technology and who is harmed.

In the absence of binding, globally enforceable regulation, companies developing GenAI often position themselves as the primary arbiters of safe deployment \cite{khanal2025bigtech}. Corporate safety narratives thus frequently serve as a de facto framework for policymakers, researchers, and the public \cite{khanal2025bigtech}. For example, OpenAI CEO Sam Altman recently testified that \textit{``the benefits of the tools we have deployed so far vastly outweigh the risks''} \cite{kang2023altman} as he urged the US Congress to regulate AI. Similar rhetoric exists in industry-developed safety standards, policy white papers, and funding of research and advisory institutions focused on AI risks \cite{ahmed2024field}. While these initiatives indicate a commitment to safety, they also reveal a central tension of self-regulation: the same actors driving GenAI’s rapid proliferation are simultaneously defining its limits and acceptable practices.

Corporate framing also shapes the AI safety research community, which frames its work as urgent and morally imperative. Yet, its narrow focus and deference to corporate interests risk limit the field’s rigor, inclusivity, and societal relevance, as responses to these tensions shape the trajectory of AI safety research \cite{ahmed2024field}. In this context, safety becomes less a collective obligation and more a strategic discourse shaped by actors with the resources to steer public understanding and regulatory agendas \cite{stardust2023surveillance, moller2006safety, 10.1145/3313831.3376167}. Analyzing corporate narratives can therefore reveal how safety is constructed and why it is framed in particular ways.

Ahmed et al. \cite{ahmed2024field} argue that institutions do not merely describe conditions but actively orient subjects toward particular norms, futures, and attachments by repeatedly aligning them with feelings of comfort, reassurance, and legitimacy. Building on this, our work examines how ``safety'' is discursively constructed and mobilized in corporate communications. Drawing on feminist literature that views safety as dynamic and intersubjective---shaped by inclusion, care, transparency, accountability, epistemic uncertainty, and structural injustice \cite{stardust2023surveillance, moller2006safety}---we examine how safety is positioned to reveal the power relations embedded in GenAI governance \cite{10.1145/3313831.3376167,harding2025what}.

This paper asks: \textit{How do GenAI companies construct the notion of safety in their public communications?} Using critical discourse analyses of safety-related documents (n=75) from three major GenAI companies, we examine both what these texts include and how they frame safety, focusing on the power dynamics and political implications embedded in the language.

Across company documents, we find that responsibility is widely distributed across companies, users, and governments while accountability remains unclear; governance is framed through internal committees and selective calls for regulation; risk is presented as expansive and urgent, requiring continuous monitoring, red-teaming, and iterative deployment. These discourses position companies as proactive stewards of AI while simultaneously limiting external oversight. By treating corporate communication itself as a key site of analysis, our work contributes to and extends HCI research on AI literacy and governance, providing a critical, discursive lens through which potential avenues for HCI intervention may be identified.

\section{Background}

Before examining AI safety, it is important to discuss the broader concept of safety and what it means to be safe, both online and offline. In psychology, safety is the second level in Maslow's hierarchy of needs, representing the human desire for security, stability, and protection from harm; some aspects of safety in this context include physical safety, health, well-being, and financial security~\cite{maslow1943theory}. While Maslow discussed \textit{safety} and \textit{security} together,~\citet{strohmayer2022safety} drew a distinction: \textit{``Security is defined as the protection from deliberate threats (such as an adversary) while safety as the condition of being protected from situations that are likely to cause harm (such as toxic workplace environments) [..] Put another way, securing known harms may not produce the protection necessary to keep a person safe from bodily, emotional, and psychological harm.''}

Prior work has documented how groups already marginalized by discrimination or oppression, such as sex workers~\cite{soneji2024feel,mcdonald2021s}, gig workers~\cite{rivera2024safer}, and LGBTQ+ communities~\cite{mcclearn2023othered,scheuerman2018safe}, experience heightened safety harms online and offline. In particular, criminologists Harris and Woodlock coined the term safety work to describe the mental and physical labor required to remain safe, such as when survivors of domestic violence navigate digital coercive control~\cite{harris2019digital}. A feminist orientation to safety, therefore, centers the needs of marginalized communities, emphasizing that technical fixes alone are insufficient; safety must also confront power structures across wider social ecologies~\cite{strohmayer2022safety}. Building on \citet{coles2020too}, post-digital perspectives similarly highlight how safety and harm are experienced across entangled digital and non-digital domains, unfolding within individuals, groups, networks, and platforms~\cite{rivera2024safer,setty2024post,campaioli2025now}. Together, these approaches underscore the need for a holistic understanding of safety that integrates technical, social, and political dimensions.

\section{Related Work}

\subsection{Conceptualizing AI Safety}

The primary concern in AI safety lies in how current and future AI models pose risks to individuals, society, and humanity as a whole, particularly as generative AI (GenAI)-based chatbots become widely adopted and repurposed. \citet{ferri2023risk} differentiate between AI safety work aimed at mitigating harms that existing AI models pose to society and research addressing the long-term risks associated with the potential emergence of artificial general intelligence (AGI)- human-level models that could produce existential, catastrophic outcomes~\cite{mclean2023risks}. Prior work has proposed numerous frameworks and taxonomies to distinguish between different AI safety risks and relevant mitigation strategies. For instance,~\citet{abercrombie2024collaborative} and~\citet{shelby2023sociotechnical} have identified several types of risks in current AI systems, including representational and allocative harms. Rauh et al.~\cite{rauh2024gaps} highlighted gaps in existing mitigation strategies, such as modality limitations (lack of evaluation for non-text modalities), risk coverage (insufficient attention to ethical and social risks), and contextual understanding (failure to capture the broader contexts in which AI systems operate). From a policy perspective, the United Nations has similarly identified ways in which AI can pose risks to human rights~\cite{untaxonomy}.

Extending the distinction between current and long-term risks, Weidinger et al.~\cite{weidinger2022taxonomy,weidinger2023sociotechnical} developed taxonomies for evaluating both immediate and systemic harms from large language models, while~\citet{uuk2024taxonomy} and Hagendorff~\cite{hagendorff2024mapping} focused on systemic risks and behaviors such as deceptive actions, power-seeking, and self-replication. Critics have argued that the emphasis on long-term, existential risks can distract from more immediate harms, including misinformation~\cite{kay2024epistemic}, perpetuation of existing inequalities~\cite{kong2025point,zhang2024structural}, and environmental impacts~\cite{hogan2024ai}. For instance, Coldicutt~\cite{coldicutt2024ai} has shown how catastrophic framings draw on cultural myths and superhero allegories to dramatize AI risk, while Helfrich~\cite{helfrich2024harms} has argued these metaphors deflect scrutiny from immediate harms. Lazar and Nelson~\cite{lazar2023ai} have criticized this long-term framing, calling instead for sociotechnical approaches that foreground accountability, inclusion, and equity.

While technical measures remain important, most AI safety work has traditionally been framed as a technical problem~\cite{eriksson2025can}. This narrow framing can obscure the social and political dimensions of safety, particularly given scholars’ arguments that safety is inherently sociotechnical \cite{AvenYlonen2018}. In this light, AI safety discourses are significant because discourses themselves constitute and exercise power \cite{Miller1990}. In particular, general-purpose AI companies produce and circulate their own technical narratives of safety through social media and other channels, shaping research priorities~\cite{OpenAI2025}. For instance, OpenAI documented a Teen AI Literacy Blueprint, emphasizing literacy as a key component of promoting AI safety~\cite{openai2025teen}. Building on this perspective, our paper examines how mainstream GenAI companies construct, articulate, and circulate particular understandings of safety, and how these framings define what is considered legitimate, necessary, and actionable.

Taking a complementary approach, Ahmed et al.~\cite{ahmed2024field} examined the emergence of AI safety as an epistemic field, focusing on how authority, legitimacy, and research priorities are shaped through academic institutions, funding streams, student groups, and competitions. Their analysis emphasized the field-building dynamics of AI safety and the risk of epistemic monoculture. In contrast, our work shifts attention from institutional organization to the discursive outputs of the most influential actors. We analyze how mainstream GenAI companies publicly articulate and frame ``AI safety'' in their materials, highlighting how these narratives construct particular ideas about the same.

\subsection{The Sociotechnicality of GenAI}

Extensive work has examined how to mitigate AI risks through several technical interventions, such as benchmarking~\cite{eriksson2025can} and different ways of training AI~\cite{mitchell2021algorithmic,pessach2022review}. Prior work demonstrates how relying solely on technical solutions for safety concerns is insufficient to address real-world concerns~\cite{10.1145/3531146.3533158, zhang2024structural,jorgensen2023not,lee2024algorithms, amsterdamfairai, kong2025point}. The biggest concern is that technical interventions largely fail to acknowledge that AI technologies are embedded in social contexts~\cite{selbst2019fairness}, often perpetuating existing societal harms and creating new risks. Framing these safety risks as purely technical problems can deflect accountability away from those developing GenAI~\cite{edenberg2023disambiguating}. 


A sociotechnical approach to AI acknowledges that the technology is mutually shaped by social actors and institutions~\cite{kong2025point,theus2023striving,birhane2021algorithmic,ehsan2022algorithmic}. Thus, any attempts to mitigate these risks must account for the fact that AI is embedded in society and existing power structures~\cite{selbst2019fairness,ovalle2023factoring,zhang2024structural,kong2022intersectionally}, while seeking to create \textit{``spaces for the wider participation [...] to steer our technological future on the basis of equal concern and common humanity''}~\cite{lazar2023ai}. To model these safety risks in a way that accounts for social structures, researchers have emphasized the need for interdisciplinary collaboration in AI research and development~\cite{selbst2019fairness,theus2023striving}. Going beyond academic circles, \citet{edenberg2023disambiguating} have called for more participatory design efforts, including those who are disproportionally impacted by AI technologies, as a way to balance technological affordance and the social needs of users~\cite{ehsan2023charting, 10.1145/3706598.3713165, 10.1145/3706598.3713510}.

GenAI is also a sociotechnical artifact; its development and interpretation are shaped by broader cultural, political, and economic forces. These forces inform sociotechnical imaginaries \cite{doi:10.1177/20594364231196547, 10.1145/3613905.3650947}---namely, shared, socially supported visions of desirable futures \cite{Dishon2024}---that frame GenAI either as a transformative tool for progress or as a source of existential societal risk~\cite{doi:10.1177/09636625251328518}. For instance,~\citet{doi:10.1177/20594364231196547} has examined how national narratives about GenAI project societal priorities and anxieties onto technology. Another way in which GenAI represents a complex sociotechnical phenomenon is through its platformization. Similar to research on social media~\cite{poell2019platformisation, nieborg2022platforms}, GenAI can go beyond its original functions to shape broader social, cultural, and economic practices. Scholars have argued that GenAI platforms are inherently transactional and marginalizing~\cite{10.1145/3419249.3420167}. By relying on internal, iterative, and proprietary processes to manage AI safety risks~\cite{gillespie2024ai}, GenAI concentrates power in the hands of the few who own, design, and produce these systems while motivated by capitalistic interests~\cite{khanal2025bigtech, Verdegem2022}. Likewise, when discussing and talking about safety in these scenarios, taking a sociotechnical lens becomes important \cite{10.1145/3715275.3732062}.

The acknowledgment of GenAI safety as a sociotechnical problem is also present in HCI literature. Recent work has surveyed the ways in which humans can interact with GenAI~\cite{shi2023hci, 10.1145/3613905.3650947} and explored how to design AI in a way that accounts for those who use GenAI to promote appropriate trust~\cite{kim2024m,kim2025fostering,mehrotra2024systematic}. A workshop held at CHI 2025 also explored how to address AI governance through a sociotechnical lens~\cite{feng2025sociotechnical}. Extending this perspective, prior work in HCI has explored toolkits as a way to understand and explore AI systems \cite{10.1145/3579621,hollanek2024ethico,10.1145/3442188.3445938}.  Approaches like Human-Centered Explainable AI (HCXAI) exemplify these perspectives by highlighting how mismatches between designer intentions and user perceptions can lead to social misattributions, thereby reinforcing the importance of accounting for social and organizational context in AI safety~\citep{ferrario2024hcxai}.

In this paper, we adopt a sociotechnical lens on GenAI, analyzing corporate safety narratives to examine how power, authority, and accountability are constructed. Our discourse analysis reveals how safety is framed, which harms are emphasized or downplayed, and how participation and governance are rhetorically portrayed.

\subsection{Studying Discourses}

In this paper, we critically examine the \emph{discourses} employed by GenAI platforms. Discourses are significant because they operate across multiple contexts, shaping meaning in ways that allow ideas to materialize as objects \cite{https://doi.org/10.1111/joms.12113}. Discourses are performative, shaping how knowledge and ideas are constructed, communicated, and normalized \cite{10.1145/3706598.3714073, doi:10.1177/0263276415571940, reinelt2002politics, vandijk2017discourse}. In technology contexts, the dominant discourses reflect existing power structures and influence whose perspectives are valued \cite{10.1145/3290605.3300569, 10.1145/1394445.1394446}. Repeated narratives contribute to the stabilization of knowledge, underscoring the importance of critically examining corporate and scholarly texts to understand how authority, legitimacy, and norms are produced \cite{jager2014analysing}.

Discourses of safety have been studied in the context of social media, during the period when companies were expanding, developing tools and features, and implementing ideals that were discursively critiqued as unsafe. Consequently, social media companies began producing public-facing safety pages and resources meant to communicate their commitment to user safety \cite{berkovich2023history}. Since then, scholars from various disciplinary perspectives and epistemologies have analyzed these pages, finding that the concept of \textit{"safety"} is often entangled with, and obscured by, capitalistic logics of surveillance, violence, and profiteering \cite{https://doi.org/10.1002/poi3.290, doi:10.1177/20563051221144315}. Similarly, prior work has also shown how corporate discourses around privacy align with business logics, producing deterministic claims about how privacy should be defined and enforced \cite{10.1145/3479565}. Research has further highlighted work on the discursive mediation of tensions between free speech and safety, often drawing on similar language but yielding divergent policies \cite{doi:10.1177/2056305119832588}. Across both current and past social media platforms, such discourses normalize evolving practices, while the labor of being and working on these systems remains precarious due to the persistent absence of standard regulations around safety and care \cite{doi:10.1177/08969205241306300}.

We build upon this scholarship by examining the discursive techniques that GenAI companies employ in their public-facing communications. We adopt a broad, context-sensitive understanding of safety following~\citet{stardust2023surveillance} and conceptualize it not merely as the absence of harm, danger, or risk but as the intertwining of inclusion, equitable access, transparency, and accountability. This approach recognizes that safety is socially and politically situated, shaped by structural and systemic factors, including the practices and policies of platforms~\cite{bartolo_matamorosfernandez_2023_online}. By framing AI safety in this way, we emphasize the importance of its sociotechnicality.

We focus on GenAI discourses due to the recognition that evolving societal narratives around AI are shaped by cultural, political, and ethical assumptions, reflecting and obscuring the complex, value-laden, and power-infused nature of AI technologies \cite{10.1145/3715275.3732083}. More specifically for AI safety, the power of discourses is evident in debates concerning how long-term, existential narratives of risk have the potential to divert attention away from more pressing---and most importantly, existing---harms posed by AI~\cite{ahmed2024field,coldicutt2024ai,helfrich2024harms,hogan2024ai}. One study exploring these narratives showed that regulatory and industry actors often prioritize narrow technical fixes (such as data bias or transparency), while sidelining deeper concerns about systemic safety, power asymmetries, and the political implications of AI deployment~\cite{10.1145/3630106.3659013}.

We situate our research within critiques of Big Tech’s influence, noting that public AI discourse is often shaped by corporate narratives \cite{monsees2023transversal, winkel2025controlling, doi:10.1177/29768624251408212}.  Ahmed et al. \cite{ahmed2024field} highlight that authority in AI safety is shaped through institutional and discursive practices such as publications, labs, funding, and prizes, which prioritize certain risks and methods while marginalizing others. We extend this perspective to corporate contexts, examining how companies use discourse to assert authority and influence perceptions of legitimate AI safety concerns. Our work contributes to the critical HCI community interested in the sensemaking, design, policy, and governance of AI. By centering discourses---which are key to understanding how power and legitimacy shape ideas that become normative in innovations---our work highlights the sociotechnical imaginaries that guide GenAI's development and deployment.

\section{Methods}


As one of the first empirical studies examining how safety is discursively constructed in GenAI platforms by corporate actors, our analysis focused on a small but diverse set of AI companies offering chatbot-based tools. We chose chatbots' parent companies because of their broad conversational utility rather than domain-specific functions \cite{trofymenko2021taxonomy}. Our goal was to develop a formative understanding of how safety is publicly framed across widely used GenAI platforms and to analyze discourses situated in real-world contexts, where users might seek information about AI safety. 



We employed critical discourse analyses (CDA) as our analytical approach, grounded in an interpretivist orientation \cite{powers2007philosophical}. CDA is particularly useful for unpacking how language constructs and legitimizes particular social realities \cite{jager2014analysing}. It allows for the explicit identification and understanding of discursive practices, particularly by showing how power is exerted and maintained through knowledge construction \cite{fairclough2013critical}. It also enables deeper and broader insights into the ``why'' and ``how'' when interrogating systems that reinforce and obscure power relations, in ways that other methods may not fully engage with~\cite{doi:10.1177/1461444816677532}. Finally, CDA helps establish how certain practices and identities become normalized or dominant, often without necessary critical examination \cite{fairclough2013critical}. In this case, it allowed us to analyze how companies developing AI tools frame safety in their public-facing materials. Below, we describe our data collection and analysis in more detail.

\subsection{Data Collection}




For selecting companies, we referred to market share data from StatCounter\footnote{\url{https://gs.statcounter.com/ai-chatbot-market-share}} and Statista\footnote{\url{https://www.statista.com/statistics/1618020/ai-chatbots-traffic-share-ww/}}. Based on this information, we selected three prominent companies with widely used GenAI based chatbots: OpenAI (ChatGPT), Google (Gemini), and Anthropic (Claude).\footnote{Although we initially considered Microsoft, our early analyses revealed that most of their documents focused on business-to-business (B2B) communications, so we decided not to include them. We also did not select Perplexity, as it is a generative search engine rather than a conversational agent unlike the others.} 

Our initial focus was on materials that explicitly mentioned "safety," but broader reading and team deliberation revealed that related concerns—such as trust, responsibility, and harm, were pervasive in the discourse. 
Scholars have similarly argued for a broader definition of safety that goes beyond the mere absence of harm \cite{stardust2023surveillance}. Following this perspective, safety can be understood as entangled with values such as inclusion, equitable access, transparency, and accountability, and as socially and politically situated, shaped by structural and systemic conditions, including platform practices and policies \cite{bartolo_matamorosfernandez_2023_online}. This understanding guided our decision to include materials beyond explicit mentions of "safety," since AI safety is multidimensional, encompassing not only formal policies but also related practices, governance decisions, and indirect mechanisms through which companies manage risk and influence outcomes.

Building on prior work on safety discourse \cite{doi:10.1177/20563051221144315}, we focused on seven keywords: harm, responsibility, trust, safety, transparency, accountability, and mitigation. Using Google Advanced Search in incognito mode, we applied the query: \textit{site:<company domain> harm OR trust OR safety OR transparency OR accountability OR mitigation OR responsibility}. While not exhaustive, these keywords generally captured how AI platforms frame safety: ``safety'' and ``harm'' refer to preventing abuse or misuse of AI, ``trust'' reflects user confidence in AI behavior, ``responsibility'' and ``accountability'' highlight the platform’s obligations to manage risks, ``transparency'' signals how policies and interventions are communicated, and ``mitigation'' points to concrete strategies such as moderation, verification, or model adjustments. OpenAI’s Transparency Report illustrates an example of this framing, explicitly linking evolving policies, monitoring of misuse, and intervention strategies to safety:

\begin{quote}
OpenAI’s Transparency Report (the “Report”), provided in accordance with the EU Digital Services Act (“DSA”). The Report includes data for content, users and reporters, as applicable, from EU member states, covering the period from February 17, 2024 through December 31, 2024 (the “reporting period”). Our policies and practices continue to evolve in conjunction with our services themselves, as well as environmental factors and patterns of potential abuse. We welcome the opportunity to discuss such changes through future reports." (10)
\end{quote}

For each company, we collected the top 25 search results. These documents were manually reviewed and filtered for relevance. We included all publicly available documents from company domains that addressed (with or without explicit mention) at least one of the safety-related keywords, without limiting by document type or model--- which meant that these included opinions about broader AI safety, or ideas about their own AI models and platforms. We excluded duplicates, broken pages, content irrelevant to safety (e.g., marketing or unrelated technical updates), and externally authored material, ensuring the corpus reflected official, institutionally endorsed discourse on AI safety. 

In total, our corpus included 38 \emph{Updates} (blog-style posts on policy changes, committee formations, other safety measures), 9 \emph{Help \& Research} documents (guidance and research-oriented materials), and 28 \emph{Perspective} pieces (interpretive or opinion-driven texts shaping broader narrative and discourse). While we cannot state with certainty the intended audiences for these materials, they are official documents affiliated with the respective companies, meaning that users, developers, journalists, and policymakers could regard them as authoritative. For pages with dynamic content, additional screenshots were captured. All data collection was completed by July 2025. The final set of documents was compiled in a spreadsheet and imported into Atlas.ti for qualitative coding and analysis.\footnote{The full dataset can be found here: https://tinyurl.com/y6dp5n56 All primary source materials are stored in the OSF project’s dataset/ folder and consist of original corporate safety documents, preserved in their original format.}

We focused on company-authored documents from their official websites because they represent the most stable, public, and institutionally endorsed articulations of corporate positions \cite{birch2022bigtech}. Using the company domain address in our search ensured that all captured documents originated from official company domains, maintaining both inclusivity of relevant materials and analytic consistency. This approach avoided including individually authored or external commentary that may not reflect the company’s institutional stance, while allowing us to systematically analyze curated outputs such as blog posts, research updates, press releases, and policy statements. Following prior research that similarly analyzed policy and governance documents to trace corporate discourse and value articulation~\cite{https://doi.org/10.1002/poi3.290, doi:10.1177/20563051221144315, 10.1145/3479565}, we limited our corpus to these materials to ensure analytic consistency, and traceability of claims within official communication channels.



\subsection{Data Analysis}

We followed \citet{jager2014analysing} in conducting CDA. Our approach combined structural, detailed, and synoptic analyses to examine how companies framed safety in their public-facing documents.\footnote{The codebook can be found here: https://tinyurl.com/y6dp5n56}

First, we conducted an iterative structural analysis to examine how the texts were organized, including the conceptualization of safety, particularly in terms of the content discussed when companies claimed to address safety. Selected examples of structural codes include \textit{Consequence of Unsafe Usage}, \textit{Developer Responsibility}, \textit{Enforcement Limitations}, and \textit{Mitigation Measures}. The first author conducted an initial reading of a randomly selected subset of safety-related documents and developed a preliminary set of structural codes. These codes were then discussed and refined with the second author, producing a revised codebook. Both authors independently coded five additional randomly selected documents per platform, after which the codebook was further refined for clarity and consistency; this process was repeated twice. The first author then used the finalized codebook to code the full dataset using Atlas.ti, ensuring a systematic and collaborative analysis of all documents.

Building on the structural analysis, the detailed analysis focused on language and discursive strategies. During initial coding, additional codes were generated to capture recurring patterns of discursive positions and framing strategies beyond structure or content, capturing how platforms established and legitimized safety. The first author discussed and refined these codes with the second author. Using this refined set of codes, the first author conducted a comprehensive analysis of the entire dataset. Selected examples include \textit{Cautious Innovation}, \textit{Metaphors}, and \textit{Dynamicity}.

Finally, a synoptic analysis was conducted to integrate and compare insights from earlier stages of the study \cite{jager2014analysing}. For this process, we reviewed the structural codes to understand the overall organization of safety-related discourse across the company-authored documents. Then, we examined the detailed codes to identify specific themes, patterns, and nuances in how safety, trust, responsibility, harm, transparency, accountability, and mitigation were articulated. During this process, we compared patterns across codes and documents, paying particular attention to both explicit statements (what was said) and implicit meanings (how it was said), including the reasons behind explicit statements, possible alternative interpretations, omissions, framing strategies, and the positioning of actors within the discourse. Finally, the insights from this review were synthesized collaboratively among the authors to reconstruct the broader assumptions and knowledge embedded in the corporate discourse, highlighting how companies conceptualized and communicated safety within broader institutional, social, and organizational contexts, and implications for HCI and other fields.

\section{Findings}

We organize our findings in two separate---yet interrelated---parts (see Table \ref{tab:findings_summary}). The first part draws on interpretive analyses of our structural codes and examines how GenAI companies structure safety in their discourse (\S\ref{sec:what_safety}), focusing on \emph{what} they discuss when addressing safety. The second part draws on analyses of our detailed codes and explores \emph{how} GenAI platforms discuss safety, highlighting the strategies they employ to shape and control the AI safety discourse (\S\ref{sec:how_safety}).

\begin{table}[t]
\centering
\caption{Summary of Findings on GenAI Companies’ AI Safety Discourse}
\label{tab:findings_summary}
\begin{tabular}{p{0.32\columnwidth} p{0.62\columnwidth}}
\toprule
\textbf{Finding Dimension} & \textbf{Brief Overview} \\
\midrule
\multicolumn{2}{l}{\textbf{\emph{What} Companies Talk About When They Talk About Safety}} \\
\addlinespace[0.4em]

Responsibility and Accountability
& AI safety is framed as a shared responsibility across companies, users, governments, and civil society, emphasizing proactive commitments and technical safeguards while leaving accountability for concrete harms weakly specified. \\[0.4em]

Governance, Oversight, and Control
& Safety is tied to internal governance structures and selective regulatory collaboration, with companies advocating for “surgical” oversight that minimizes external constraints while preserving innovation. \\[0.4em]

Risk, Uncertainty, and Harm Mitigation
& Companies enumerate a broad range of risks---from bias and misuse to catastrophic and existential threats—and emphasize continuous evaluation, red-teaming, monitoring, and iterative mitigation practices. \\
\midrule
\multicolumn{2}{l}{\textbf{\emph{How} Companies Talk About AI Safety}} \\
\addlinespace[0.4em]

Constructing Authority
& Companies establish legitimacy through claims of technical expertise, ethical positioning, and expert-led governance structures, presenting themselves as indispensable stewards of AI safety. \\[0.4em]

Dynamic and Global Framing
& AI safety is framed as an evolving, iterative process that cannot be fully resolved pre-deployment and as a global concern requiring multi-stakeholder coordination. \\[0.4em]

Metaphors and Analogies
& Safety is articulated through metaphors drawn from high-risk domains (e.g., nuclear power, aviation, CBRN threats) and operational analogies that normalize uncertainty while legitimizing ongoing oversight. \\
\bottomrule
\end{tabular}
\end{table}

\subsection{\emph{What} Do Companies Talk About When They Talk About Safety?}
\label{sec:what_safety}

In tracing what companies mean when they invoke ``safety,'' our analysis identified three major dimensions: 1) responsibility and accountability, where companies define their own roles while distributing obligations across others; 2) governance, oversight, and control, where safety is tied to internal structures and external collaborations; and 3) risk, uncertainty, and harm mitigation, where companies acknowledged unpredictability while promoting technical safeguards.

\subsubsection{Framing Responsibility and Distributing Accountability}

Our analysis of accountability and responsibility traces three interrelated dimensions: (1) companies themselves defining and enacting responsibility through technical safeguards and corporate policies; (2) the shifting of responsibility to users and other stakeholders, including governments; and (3) the implications of distributed responsibility for accountability. Together, these dimensions illustrate how companies projected an image of proactive governance while diffusing accountability for the interdependent AI ecosystem.

Across documents, we saw frequent emphasis on responsibility as central to AI safety, framing it as corporate commitments, technical safeguards, and evaluative processes. OpenAI's communication began with a high-level commitment: \textit{``OpenAI is committed to keeping powerful AI safe and broadly beneficial''} (15), going as far as to dictate \textit{Our primary fiduciary duty is to humanity} (121).

Using a similar strategy, Anthropic struck a more process-oriented tone: \textit{``We’re committed to evolving our approach alongside these developments, including adapting our frameworks, refining our assessment methods, and learning from both successes and failures along the way.''} (29)


%

OpenAI further reassured enterprise customers that \textit{``we don’t train our models on your organization’s data by default''} (4), positioning responsibility as a commitment to data-handling practices. They further noted that \textit{"like any technology, these tools come with real risks---so we work to ensure safety is built into our system"} (15), highlighting proactive responsibility-taking. This proactive---and rather technical---responsibility-taking process was also featured in OpenAI's commitment to test their models and conduct red-teaming ``before release'' (19) aligning with broader shifts in which tech companies moved from the denial of responsibility toward articulating it through ethical commitments\cite{doi:10.1177/14614448241229406}.

At the same time, companies frequently emphasized that responsibility should be distributed across multiple actors, often involving internal governance structures, authority figures, and external stakeholders. Anthropic described \textit{``independent checks''} and \textit{``assurance structures''} modeled after high-risk industries (43, 45), emphasizing that \textit{``We know we can't do this work alone. We invite researchers, policy experts, and industry partners to collaborate with us as we continue exploring these important questions''} (29). OpenAI relied on committees, preparedness frameworks, and expert advisors (3, 19). Google highlighted the need for \textit{``concrete, context-specific guidance from governments and civil society''} (63), but noted that \textit{``ultimately it is companies and developers who are at the frontline of defense from bad actors''} (63). 


Responsibility also extended to users, who were expected to adapt to the perceived increasingly capable AI. Companies discussed the importance of caring for \textit{``psychological factors''} (63), such as automation bias and algorithm aversion, to ensure users do not \textit{``place more faith in its correctness than is warranted''} (63) or \textit{``ignore safety-critical guidance''} (63). Policies, such as Google’s Generative AI Prohibited Use Policy (68), explicitly outline how users should engage with AI responsibly, transferring part of the safety burden to end-users. Reporting mechanisms further encouraged user participation in mitigating AI risks (21).

Governments and other public actors were often explicitly mentioned when discussing responsibility, particularly in the context of AI deployment within society. OpenAI stressed collaboration between \textit{``policymakers and AI providers [to] ensure that AI development and deployment is governed effectively at a global scale''} (15), and Anthropic urged governments to act promptly on AI policy (47). Similarly, Google acknowledged that governments and civil society ultimately shape the frameworks within which AI systems are deployed (63). These calls positioned public actors as essential to maintaining safe and responsible AI.

On the other hand, accountability, defined as the obligation to explain and justify the actual harms caused by GenAI models ~\cite{bovens2007analysing}, remained ambiguously defined. That is in contrast to forward-looking responsibility, which companies portrayed as a shared duty among firms, regulators, users, and civil society. In that context, Google described,

\begin{quote}
\textit{``Lead on and help shape responsible governance, accountability, and regulation that encourages innovation and maximizes the benefits of AI while mitigating risks (e.g., our role in setting up Partnership on AI, our support for Global Partnership on Artificial Intelligence and our contributions to flagship AI governance efforts, including the EU AI Act, NIST AI Risk Management Framework, and OECD AI Principles).''} (58)
\end{quote}

While this emphasized proactive measures, governance, and risk mitigation, it did not clarify who would actually face consequences when harms occur. By framing responsibility separately from enforceable accountability, companies presented themselves as diligent while leaving the question of real-world consequences largely unresolved. 


\subsubsection{Governance, Oversight, and Control}

The discourse around AI safety was also structured through the promotion of internal and external governance structures, which combined public-facing policies with private coordination. By calling for partnerships and participatory approaches to safety, AI companies signaled openness in recognizing that \textit{``getting AI right requires a collective effort''} (58). Furthermore, companies showed support for independent assessments as a way to address limitations in their own AI evaluations.

Companies developed its own internal governance structures for AI safety. These structures took the form of dedicated teams, such as OpenAI's Safety and Security Committee (3), and formal policies for managing risk, like Anthropic's \textit{``clearly-articulated policy on catastrophic risks'' (43)}. Companies often defined specific public-facing criteria and multi-stage testing processes to evaluate whether a particular AI model was safe, as exemplified by Anthropic's \textit{``Red Line Capabilities''} framework and AI Safety Levels (43).

Although companies promoted their own standards, they positioned themselves as collaborators with governments and policymakers. They framed enforceable regulations, including public transparency requirements for risk policies and evaluation results, as necessary to build trust and ensure accountability. This discourse emphasized the need for a collective, multi-stakeholder approach to effective regulation.

Importantly, the documents also emphasized that regulations should be \textit{``surgical''} and \textit{``simple to understand''} (47) to avoid stifling innovation. Companies were openly cautious about how much external stakeholders could mediate, stating: 

\begin{quote}
\textit{``Overall, Google recommends a cautious approach for governments regarding liability for AI systems, since the wrong frameworks might place unfair blame, stifle innovation, or even reduce safety. Any changes to the general liability framework should come only after thorough research establishing the failure of the existing contract, tort, and other laws.''}  (63)
\end{quote}

Google employed notably assertive language, explicitly recommending a pro-innovation approach in commentary on the U.S. AI Action Plan (126).Likewise, Anthropic, communicated a similar idea: 

\begin{quote}
     \textit{``Whatever regulations we arrive at should be as surgical as possible. They must not impose burdens that are unnecessary or unrelated to the issues at hand. One of the worst things that could happen to the cause of catastrophic risk prevention is a link forming between regulation that’s needed to prevent risks and burdensome or illogical rules.''} (47)
\end{quote}

This call for some kind of external governance structure also included independent third-party evaluations, where companies advocated to inform \textit{"new standards and laws"} (19), or "\textit{to evaluate the effectiveness of our [their] security controls}" (48), signaling lightly that external oversight is essential to managing AI safety risks.

In summary, the documents showed how their internal teams and safety frameworks were core to their business functions. Likewise, we also noted comapnies' willingness to collaborate with governments and policymakers to shape regulations in pro-innovation ways that were tailored to their specific concerns.


\subsubsection{Managing Risks and Mitigation Practices}

AI safety was also operationalized through processes and evaluation metrics that identified risks and imposed constraints. Companies framed their efforts as part of a continuous cycle of risk management: \emph{identifying and anticipating harms, testing models at defined stages, limiting or steering their use, and iterating on these steps over time}. Within this cycle, transparency, data sharing, and user feedback were positioned as feedback mechanisms that reinforced safety. In effect, \textit{``safety''} was constructed as an ongoing process of proactive identification and mitigation of risks.

Companies frequently identified a wide spectrum of AI-related risks, including complex and potentially catastrophic dangers beyond familiar concerns such as toxicity and bias, often implying which harms to prioritize. Anthropic, for instance, highlighted that AI behaviors may diverge from designer intentions, including \textit{``sycophancy and a stated desire for power''} (35). Google cataloged risks ranging from technical failures to societal harms such as the amplification of biases and the creation of \textit{``information hazards''} through misinformation and nonfactuality (58). These acknowledgments further encompassed existential threats, including \textit{``large-scale devastation through deliberate misuse''} by \textit{``terrorists or state actors to create bioweapons''} or autonomous actions \textit{``contrary to the intent of their designers''} (44), with Google emphasizing the risk of deceptive alignment, in which systems recognize conflicts with human instructions and deliberately circumvent safeguards (116).


In response to these (and also other) concerns, companies emphasized systematic evaluation and testing as central to their mitigation strategies. Anthropic explicitly argued that \textit{``any industry where there are potential harms needs evaluations,''} drawing analogies to nuclear power monitoring and aircraft flight testing (45). Methods included red-teaming to stress-test models and creating defenses against adversarial misuse. Anthropic advertised its \textit{``Constitutional Classifiers''}, which were designed to guard against \textit{``universal jailbreaks,''} i.e., attempts to elicit forbidden responses (34). Complementing these defenses, companies deployed monitoring systems to detect subtle patterns of misuse. Anthropic’s Clio system, for example, enabled \textit{``privacy-preserving analysis of real-world language model use''} to uncover \textit{``coordinated, sophisticated misuse''} that would be invisible at the level of individual interactions (33).

The discourse of risk mitigation was reinforced through organizational and regulatory commitments, as discussed above. OpenAI formed a dedicated safety committee \textit{``responsible for making recommendations on critical safety and security decisions''} (3), while Anthropic created a \textit{``Responsible Scaling Team''} (43). Google emphasized enforcement through its \textit{``Generative AI Prohibited Use Policy''} (68), and all companies pointed to compliance with governmental regulatory frameworks, such as the European Union's GDPR and the US HIPAA (4). Transparency reports detailing banned accounts, government requests, and key metrics were presented as evidence of responsibility and industry leadership.

The last step of the \textit{``safety cycle''} refers to the role of iterative feedback in promoting AI safety. Google emphasized that they \textit{``listen, learn and improve based on feedback from developers, users, experts, governments, and representatives of affected communities [..] and involve human raters to evaluate AI models''} (58), noting that \textit{``human users provide essential feedback to improve AI systems over time''} (63). 


This process of feedback closed the loop in the safety cycle: risks were first acknowledged, investigated through evaluation and testing, countered with mitigation tools and governance structures, and continuously refined through user feedback. This iterative process allowed companies to present continuous model refinement and adaptation of safety measures, while signaling responsiveness to real-world use and evolving risks.

\subsection{\emph{How} Do Companies Talk About AI Safety?}
\label{sec:how_safety}

\subsubsection{Constructing Authoritative AI Safety}

Weber \cite{weber1922_economy, weber1919_politics} defined authority as legitimate power accepted by others, based on tradition, charisma, or legal-rational rules. In our corpus, we observe such power as discursively constructed. Authority was enacted through narratives, institutional arrangements, and public statements that demonstrated how these companies were uniquely positioned to manage both the technical and societal consequences of AI, performing legitimacy through strategic narratives and positioning--- such as OpenAI talking about its own charter: \textit{``The timeline to AGI remains uncertain, but our Charter will guide us in acting in the best interests of humanity throughout its development.''} (121)


Companies constructed authority through technical expertise, ethical positioning, and formal governance. Boards, committees, and expert networks reinforced legitimacy, while forward-looking practices like risk assessment signaled the capacity to anticipate harms. Together, these elements positioned companies as essential stewards of AI, combining expertise, ethics, and institutional oversight.

OpenAI constructed authority by framing itself as the steward of managing societal risks from advanced AI. The company stated that \textit{``to be effective at addressing AGI’s impact on society, OpenAI must be on the cutting edge of AI capabilities—policy and safety advocacy alone would be insufficient''} (121), signaling that technical expertise and proximity to the technology were prerequisites for legitimate guidance. Authority was further enacted institutionally through the formation of the Safety and Security Committee, led by Board members Bret Taylor (Chair), Adam D’Angelo, Nicole Seligman, and Sam Altman (CEO), and supported by technical and policy experts including Aleksander Madry, Lilian Weng, John Schulman, Matt Knight, and Jakub Pachocki. The committee also consulted former cybersecurity and national security officials, including retired U.S. Army General Paul M. Nakasone, Rob Joyce, and John Carlin (3). By naming prominent experts and invoking its Charter, OpenAI performed epistemic and moral authority while positioning itself as uniquely capable of anticipating and managing AI-related risks.
They further reinforced authority through long-term governance narratives, stating that \textit{``The timeline to AGI remains uncertain, but our Charter will guide us in acting in the best interests of humanity throughout its development,''} performing authority through anticipatory governance and ethical framing.

Anthropic discursively constructed authority by emphasizing deliberative governance and expert-guided decision-making. As a Public Benefit Corporation, it framed its mission around \textit{``ensuring a safe transition through transformative AI''} (40). The document claimed:

\begin{quote}
    \textit{``Anthropic's Long-Term Benefit Trust today announced the appointment of Richard Fontaine, CEO of the Center for a New American Security, as a new member of the Trust. The Long-Term Benefit Trust (LTBT) is an independent body designed to help Anthropic achieve its public benefit mission.''}
\end{quote}

Authority was further enacted through the LTBT’s guiding role in leadership decisions, as highlighted when a trustee noted, \textit{``My conversations with Dario, Daniela, and the LTBT team made clear that Anthropic takes these challenges seriously. As the stakes get higher, the LTBT serves as a valuable mechanism to help Anthropic’s leadership navigate critical decisions. I am pleased to lend my expertise to this important work''} (40). 


Google discursively constructed authority by framing itself as the central actor in AI governance while circumscribing the role of external oversight. The company argued that \textit{``self- and co-regulatory approaches remained the most effective practical way to address and prevent AI-related problems in the vast majority of instances''} (63), signaling legitimacy by framing its authority as inherently pragmatic and universally applicable, signaling control and expertise while leaving operational specifics ambiguous. Google simultaneously acknowledged extraordinary societal risks, stating that 

\begin{quote}
    \textit{``Some contentious uses of AI could have such a transformational effect on society that relying on companies alone to set standards was inappropriate---not because companies can’t be trusted [..] but because to delegate such decisions to companies would be undemocratic.''} (63)
\end{quote}

By framing these cases as exceptional, Google reinforced its authority in routine governance, presenting itself as the default steward of AI while exercising strategic control over which issues required external oversight.

\subsubsection{Dynamic and Global: Multi-Stakeholder Calls for AI Safety}

Our analyses showed that companies discursively constructed AI safety as both \textit{dynamic} and \textit{global}. These framings appeared mutually reinforcing: the inherent uncertainty of \textit{dynamic} safety provided the rationale for treating it as \textit{globally} significant, while the worldwide reach and impact of AI reinforced the necessity for safety to remain adaptive, iterative, and responsive to emergent circumstances.


Safety was framed as dynamic in multiple ways, encompassing both technical and organizational dimensions. OpenAI, for instance, reported that \textit{"after our latest model, GPT-4, finished training, we spent more than 6 months working across the organization to make it safer and more aligned prior to releasing it publicly"} (15). This description emphasized that safety work extended well beyond model training and was deeply embedded in organizational practice, requiring sustained cross-team coordination and evaluation. At the same time, the company recognized the limits of predeployment efforts, acknowledging that,

\begin{quote}
    \textit{``We work hard to prevent foreseeable risks before deployment, however, there is a limit to what we can learn in a lab. Despite extensive research and testing, we cannot predict all of the beneficial ways people will use our technology, nor all the ways people will abuse it.''} (15) 
\end{quote}

This statement positioned safety as an inherently emergent and contingent phenomenon: because AI systems interact with complex human and societal contexts in unpredictable ways, companies justified ongoing iterative practices as essential for managing uncertainty.

The iterative dimension of safety extended beyond technical refinement to include stakeholder engagement and gradual deployment strategies. OpenAI described its approach as \textit{``cautiously and gradually releas[ing] new AI systems with substantial safeguards in place to a steadily broadening group of people and make continuous improvements based on the lessons we learn''} (15). By releasing systems incrementally and learning from user interactions, the company framed iteration as a means to both mitigate unforeseen harms and generate empirical knowledge of system behavior in practice. Iteration was also explicitly linked to public participation:

\begin{quote}
\textit{``Society must have time to update and adjust to increasingly capable AI, and [...] everyone who is affected by this technology should have a significant say in how AI develops further. Iterative deployment has helped us bring various stakeholders into the conversation about the adoption of AI technology more effectively than if they hadn’t had firsthand experience with these tools.''} (15) \end{quote}


Other companies echoed this adaptive framing, further emphasizing the evolving nature of AI safety. Anthropic described its approach as \textit{``still evolving. We’re sharing our current thinking while acknowledging it will continue to develop as we learn more. We welcome collaboration from across the AI ecosystem as we work to make these systems benefit humanity''} (29). 
Anthropic also emphasized the formalization of this iterative process through its Risk and Safety Plan, noting that \textit{``We regularly measure the capabilities of our models and rethink our security and safety approaches in light of how things have developed''} (47).

Google similarly highlighted the evolving nature of AI risk and the need for continuous monitoring, stating that \textit{``AI comes with complexities and risks, and these will change over time. As an early-stage technology, its evolving capabilities and uses create potential for misapplication, misuse, and unintended or unforeseen consequences''} (58). To manage these dynamically occurring risks, Google claimed to \textit{``develop methods to monitor deployed systems, ensuring that we can quickly mitigate dynamically-occurring risks in production and in-use services''} (58). Across these examples, safety was constructed as a disciplined---yet flexible---practice, described as \textit{``principled and adaptable to keep up with the evolving AI landscape''} (29). Collectively, these statements framed safety as an ongoing process of learning, adaptation, and reflection rather than as a static goal that could be fully achieved at a single point in time.

The \textit{dynamic} framing of safety provided the rationale for presenting it as \textit{global}. Because AI systems could produce unintended consequences that extended beyond localized contexts, companies emphasized the worldwide stakes of safety management. Google's discourse explicated: 

\begin{quote}
    \textit{``Our approach to developing harnessing the potential of AI is grounded in our founding mission to organize the world’s information and make it universally accessible and useful and it is shaped by our commitment to improve the lives of as many people as possible. It is our view that AI is now, and more than ever, critical to delivering on that mission and commitment.''} (58)
\end{quote}

OpenAI, on the other hand, underscored the scope of its responsibility, noting that \textit{``more than a hundred million users and millions of developers rely on the work of our safety teams''} (19). By emphasizing the large, dispersed population affected by AI systems, the company framed safety as consequential not only for individual users but also for broader societal and technological ecosystems. Google extended this reasoning, describing responsible AI as \textit{``a collective effort [..] involving researchers, developers, users (individuals, businesses, and other organizations), governments, regulators, and citizens''} (58). 



\subsubsection{Metaphors}

Our analyses revealed that companies discursively constructed AI safety by drawing on metaphors, historical analogies, and procedural frameworks to convey its stakes. A metaphor is a way of understanding one thing in terms of another \cite{lakoff2003_metaphors}, and in this context, metaphors can reveal specific comparisons that shape baseline ways of understanding what safe AI is and how it can be operationalized \cite{kajava2023_language}. While most metaphors were not explicated in boundless ways, they implied comparisons to other technologies that created particular imaginaries.


%
First, safety was consistently constructed through comparisons to dual-use technologies and catastrophic scenarios. Anthropic, for example, highlighted the potential of AI to amplify threats in domains traditionally associated with existential risk, positioning it also as generating extreme and unpredictable harms:

\begin{quote}
\textit{``Chemical, biological, radiological and nuclear (CBRN) risks---We're prioritizing evaluations that assess two critical capabilities: a) the potential for models to significantly enhance the abilities of non-experts or experts in creating CBRN threats, and b) the capacity to design novel, more harmful CBRN threats.''} (27)
\end{quote}


OpenAI likewise underscored the temporal and strategic dimensions of oversight, framing safety as a long-term investment: \textit{``We view safety as something we have to invest in and succeed at across multiple time horizons, from aligning today’s models to the far more capable systems we expect in the future''} (19). OpenAI also described its release of ChatGPT as a \textit{``Rorschach test,''} a metaphor suggesting that safety itself was interpreted through projection: depending on whether one believed AI progress to be continuous or discontinuous, the release appeared either as a reckless gamble or as an opportunity to learn (9). Collectively, these framings constructed a landscape dominated by hypothetical threats, foregrounding the scale and unpredictability of potential harms while leaving the question of practical mitigation largely unaddressed.

The metaphorization of AI safety was reinforced through technological and historical analogies. Anthropic compared AI safety practices to established safety regimes in other high-risk industries: \textit{``Nuclear power stations have continuous radiation monitoring and regular site inspections; new aircraft undergo extensive flight tests to prove their airworthiness''} (45). By linking AI to domains with robust regulatory and inspection frameworks, companies constructed a discourse in which AI was both societally consequential and technically tractable but only under sustained, rigorous oversight. Google elaborated on this approach through references to transformative technologies of the past: \textit{``In thinking through these issues, it may be helpful to review how the world has responded to the emergence of other technologies presenting ethical (and at the extreme, existential) questions''} (63). They cited genetic engineering, in vitro fertilization, nuclear technology, PCBs, and space exploration as analogues, demonstrating how voluntary and multilateral norms had historically mitigated risks while enabling innovation (63). For instance, the Asilomar Conference on Recombinant DNA and the Outer Space Treaty were invoked to exemplify how collective international coordination could provide frameworks for responsible governance (63). These analogies also functioned rhetorically, framing companies as inheritors and interpreters of past governance successes and failures, tasked with applying these lessons to a novel, high-stakes technology.

Operational metaphors further concretized AI’s risks and the challenges of alignment. Google’s examples of misaligned agents illustrated both the technical and ethical dimensions of safety: \textit{``Suppose a cleaning robot maker set the objective to remove visible dirt as fast as possible. If the optimal approach turned out to be hiding dirt under the carpet, or throwing away all visible dirty objects, this would be a failure in spirit even though it might satisfy the objective''} (63). A similar scenario, \textit{``a robot barista tasked with delivering coffee in the shortest time possible might (if given free rein) come up with the solution to throw the cup!''} (63), highlighted the tension between optimizing objective functions and unintended harmful outcomes. These examples translated abstract alignment problems into everyday scenarios, legitimizing the need for safeguards, constrained exploration, and iterative testing. Google further connected these operational metaphors to practices for establishing standards of explanation and due diligence: \textit{``Researchers from the public, private, and academic sectors should work together to outline basic workflows and standards of documentation for specific application contexts which would be sufficient to show due diligence in carrying out safety checks (e.g., like for airline maintenance)''} (63). In other words, operational metaphors provided a bridge between theoretical risk and the practical procedures companies were using to manage it.


\section{Discussion}

Our findings illuminate both what was discussed within safety and safety-adjacent documents and how these discussions were framed. We found that responsibility, accountability, governance, and risk mitigation constituted core thematic concerns in articulations of AI safety. Furthermore, we observed an implicit form of authority construction in which corporate or corporate-state entities were positioned as the appropriate stewards of AI safety and regulation. Although corporate discourse on AI safety included outside actors, such as governments, civil society, and practitioners, the power to set priorities and the epistemic authority remained firmly within the companies. These documents also created leeway for failures by emphasizing the inherent unpredictability of AI systems. Finally, they relied heavily on technological analogies as metaphors to frame AI as both vast and difficult to control.
Within HCI, we see our work as formative and exploratory, with potential contributions to AI literacy and governance in particular. We contextualize our findings in the following sections.


\subsection{AI (Safety) Literacy}

Efforts around AI literacy within HCI have largely centered on helping users understand what AI is and how AI systems function \cite{10.1145/3313831.3376727}. Existing work has emphasized explainability \cite{10.1145/3461778.3462131, 10.1145/3613249, 10.1145/3643834.3660722} and mental models \cite{10.1145/3713043.3728856}, aiming to make machine learning systems more legible to non-expert users through transparency mechanisms, interfaces, and pedagogical tools~\cite{kim2024m,kim2025fostering}. Parallel work in explainable AI (XAI) has sought to provide post-hoc rationales for algorithmic outputs \cite{10.1145/3301275.3302316}, supporting trust calibration, debugging, and user comprehension. At the same time, in tangential academic communities, substantial effort has been dedicated to auditing \cite{gilpin2018explaining, 10.1145/3757702}, measuring, and mitigating AI harms through benchmarks, red-teaming, and evaluation frameworks \cite{10.5555/3692070.3693501, barrett2024benchmark}. Yet, users develop their understanding of AI and their risks not only through direct use, but also through company communications, media, and educational materials~\cite{MagalhaesSmit2025}, including those about AI safety---which we have systematically studied.  


Our work extends prevailing notions of AI literacy by arguing that it should move beyond technical understanding and output evaluation to encompass a critical, discursive dimension, enabling individuals to examine AI’s ethical implications, societal impact, underlying assumptions, and potential alternatives within the socio-technological system \cite{velander2024critical}. As shown in our analysis, AI companies seek to explicate their AI safety practices, position themselves as authorities, and set priorities for future AI development. Being public-facing, these documents often function as primary sites through which companies communicate their safety commitments to users, making them powerful pedagogical artifacts in themselves. Reading them critically is therefore foundational to AI safety literacy.

Importantly, this critical-discursive framing aligns with policy efforts. The European Union’s AI Act explicitly defines AI literacy as the capacity to understand potential risks and harms associated with AI \cite{EU_AI_Act_2024} and calls for promoting AI literacy and public awareness through multiple channels, including training, guidelines, and best practices for stakeholders interacting with AI systems \cite{EU_AI_Act_2024_Art66}. By complementing these regulatory objectives, a discursive approach to AI literacy emphasizes not only comprehension of risks and opportunities but also the critical evaluation of corporate narratives, rhetorical strategies, and accountability structures embedded in AI governance.

Furthermore, scholars have argued that AI literacy should be framed as a set of competencies \cite{10.1145/3685680, 10.1145/3313831.3376727} such as verifying information against trusted sources, understanding outputs, and documenting or flagging problematic system behavior. Prior HCI work has demonstrated the potential of such practices: Deng et al., \cite{10.1145/3584931.3611279} engaged stakeholders in end-user contestation of AI claims, while auditing research has examined how LLM outputs can be evaluated within educational contexts \cite{10.1145/3706598.3713714}. We build on these contributions by explicitly foregrounding the need to interrogate assumptions, omissions, power relations, and political motivations embedded in AI safety narratives.

We introduce the notion of a \textit{discursive toolkit} as a mechanism and artifact to support AI safety literacy. Prior work has shown that toolkits are artifacts that shape organizational imaginaries. For example, Wong et al. demonstrated that AI-ethics toolkits encoded specific institutional visions and influenced how practitioners interpreted responsibility \cite{10.1145/3579621}; Hollanek highlighted that diverse toolkits often served symbolic ethico-political purposes limiting their capacity for change \cite{hollanek2024ethico}; and Krafft et al. developed an AI-policy toolkit for community advocates to open up AI policy as a public problem \cite{10.1145/3442188.3445938}. However, these toolkits largely focus on designing and deploying AI tools rather than explicitly \textit{communicating} technology. We argue for toolkits that help users identify recurring metaphors, recognize patterns of responsibility diffusion, and question who is positioned as accountable for, for instance, mapping environmental and labor costs associated with prompts \cite{kumar2024taking} or reflecting on how GenAI tools shape decision-making and ethical responsibility \cite{10.1145/3478431.3499308}. By situating these practices in HCI scholarship on power and accountability \cite{10.1145/3702652.3744217, 10.1145/3544549.3573808}, AI literacy can move beyond interpreting and knowing the outputs to asking questions about the systems in use, its parent companies, its background workings and critically interrogating the narratives they produce.

Our analysis further shows that AI companies frame safety along multiple, interrelated dimensions, distributing responsibility across internal teams, external experts, users, and governments while leaving accountability for actual harms ambiguously defined. Table \ref{tab:ai_literacy} summarizes some of these dimensions with concrete examples, linking each to key literacy insights. Because our work is interpretive, the table serves as a \textit{non-exhaustive}, analytic component that could inform the development of future AI literacy toolkits. It illustrates how users could be supported in identifying accountability structures, analyzing corporate safety narratives, and reflecting on the social and ethical implications of AI. By grounding these analytic components in broader AI literacy goals, future toolkits could operationalize interpretive and critical competencies, supporting users in navigating AI safety as a dynamic, socially negotiated, and context-dependent process.



At the same time, caution is needed: GenAI companies are increasingly promoting AI literacy on their own terms,\footnote{\url{https://academy.openai.com/}} embedding products directly into classrooms and curricula \cite{openai2025teachers}. A critical form of AI literacy must therefore deliberately maintain analytical distance from corporate framings, equipping users to assess claims and institutional agendas rather than accepting them at face value.

\begin{table*}[h!]
\centering
\caption{Exploratory dimensions of AI safety discourse, showing (1) recurrent discursive dimensions identified across company safety documents, (2) representative examples of how each dimension was articulated in practice, and (3) the corresponding AI literacy insights that these articulations afford, revealing how safety is framed, operationalized, and made legible to the public.}
\label{tab:ai_literacy}
\begin{tabular}{@{}p{4cm} p{6cm} p{7cm}@{}}
\toprule
\textbf{Dimension} & \textbf{Examples} & \textbf{AI Literacy Insight} \\
\midrule
Responsibility \& Accountability & Forward-looking commitments,
distributed across companies, users, and governments; accountability often vague 
& 
Responsibility $\neq$ enforceable accountability; signals diligence without clear consequences \\
\midrule
Governance \& Oversight & Internal safety teams, policies, multi-stakeholder collaboration; cautious calls for regulation
& Recognize authority construction and regulatory shaping through organizational structures \\
\midrule
Risk \& Mitigation & Iterative evaluation, testing, red-teaming, monitoring, user feedback
& Safety as an ongoing process, not a static property \\
\midrule
Dynamic \& Global Framing & Safety portrayed as emergent, evolving, and globally consequential
& Safety is contingent, context-dependent, and socially negotiated \\
\midrule
Metaphors \& Analogies & Comparisons to CBRN, nuclear power, aviation; operationalized through everyday scenarios
& Connects abstract AI risks to familiar real-world scenarios, highlighting both constraints and potential misunderstandings of AI’s impact relative to existing technologies \\
\bottomrule
\end{tabular}
\end{table*}

\subsection{Governance in AI: Discursive Framing, Participatory Tensions, and Corporate Power}

AI governance has become a growing focus within HCI, as scholars examine how participatory design, civic engagement, and sociotechnical interventions can support more inclusive and accountable forms of governance. The CHI 2025 Sociotechnical AI Governance workshop, organized by Feng et al. \cite{feng2025sociotechnical}, exemplifies this interest. The workshop brought together policymakers, designers, technologists, and affected communities to explore participatory governance models, documentation practices, stakeholder-centered design, and cross-sector communication mechanisms. Papers featured in the workshop highlighted approaches for bridging stakeholder perspectives \cite{yokota2025stakeholder}, designing tools to enable literacy for regulators \cite{svanaes2025ai}, and fostering civic engagement in AI governance \cite{islam2025public}, all aimed at facilitating meaningful participation and collaborative governance in AI systems. Importantly, these discussions intersect with policy frameworks such as the EU AI Act, which seeks to promote AI literacy, transparency, and risk-based oversight \cite{wachter2023limitations}.

Building on this scholarly context, our findings show that the same language of participation, collaboration, and multi-stakeholder engagement can be strategically appropriated by AI companies to reinforce their own authority. Companies frequently foreground collaboration in corporate communications. For example, Anthropic states: \textit{"We welcome collaboration from across the AI ecosystem as we work to make these systems benefit humanity"} (29). Yet this rhetoric coexists with resistance to regulatory constraints, exemplified by their caution that regulations should be as surgical as possible, not burdening companies (47).

Such examples illustrate how participatory language can coexist with efforts to limit oversight. While attempts to make AI governance, design, and deployment more participatory exist~\cite{Delgado2023, Tseng2025} we argue that corporate-driven communication also functions as a form of governance, shaping public expectations while narrowing the space for accountability. By critically analyzing corporate discourse, our work highlights the risks of co-optation and symbolic participation and challenges the assumption that stakeholder engagement is inherently equitable or meaningful. The ways in which participation is framed shape how people interact with systems, make decisions, and understand their role, meaning that rhetoric itself influences both user experience and the design of participatory processes.

These dynamics resonate with longstanding critiques of self-regulation and corporate influence on policy formation. Our analysis finds that companies frequently portray themselves as self-regulators, emphasizing forward-looking responsibility and proactive management of potential harms---a discursive strategy that signals accountability without binding external oversight. In practice, this often appears through momentary and opportunistic references to \textit{governance,} reflecting what \citet{doi:10.1177/2053951720919968} described as \textit{``governance at a distance: to navigate the complexities of the present, it is deemed better to aim for a horizon that is as remote as possible and hope for the best.''} Examples include OpenAI’s emphasis on multi-horizon safety planning (19) and Anthropic’s evolving safety frameworks (29). These discourses illuminate gaps between stated commitments and operational practices, underscoring the importance of mechanisms that translate participatory ideals into actionable oversight because self-regulation and forward-looking responsibilities rarely produce genuine accountability \cite{doi:10.1177/17470161231220946}.

 Moreover, even when formal regulations exist, such as EU AI Act and the revised EU Product Liability Directive, companies can dilute their impact \cite{wachter2023limitations} or slow down their implementation \cite{theverge2025scaleback}. Furthermore, corporate influence extends to shaping policy narratives. EU leaders have repeated corporate talking points, while US regulators echo long-term concerns about AI risks, aligning with industry framing \cite{marantz2024ai}. These examples highlight how companies not only resist regulation but also actively shape regulatory discourse to maintain operational flexibility. Lawsuits over data use and authorship \cite{getty2023stability,teen2025lawsuit} and lobbying around AI regulation demonstrate that voluntary corporate transparency can be incomplete or strategically deployed. Internal governance structures, safety committees, and epistemic authority-building events \cite{ahmed2024field} further illustrate how corporations consolidate control over both AI systems and public narratives about their safety.

Companies also rely on normatively resonant metaphors to communicate risk, safety, and responsibility, constructing and stabilizing organizational culture \cite{ForsterSkop2025}. GenAI systems are described as "frontier technologies," safety processes likened to ``aircraft stress-testing'' or ``nuclear incident monitoring,'' and misalignment compared to ``containment breaches'' or ``hazardous-material spills.'' These metaphors draw on historically familiar narratives of scientific risk, military defense, and crisis management, legitimizing corporate authority while underscoring the stakes involved in AI safety. These linguistic and cultural strategies reinforce the broader patterns of selective engagement with governance, connecting back to the earlier discussion of symbolic participation and corporate framing.

This aligns with Ahmed et al.’s concept of an \textit{epistemic culture} of AI safety \cite{ahmed2024field}, in which authority is produced through interconnected networks of institutes, research grants, media, competitions, policy engagement, and career infrastructures. These ecosystems consolidate power within a relatively narrow group of actors, creating an epistemic monoculture in which certain forms of risk, safety, and legitimacy dominate. Sociotechnical imaginaries further illuminate this process \cite{10.1145/2675133.2675229}: by framing the future of AI as inherently dependent on corporate stewardship, these imaginaries guide both policy and design decisions, embedding assumptions about stakeholders, public benefit, and technological trajectories. Our analysis also resonates with HCI scholarship on \textit{futuring}, showing how speculative visions influence what becomes thinkable, fundable, and legitimate \cite{Sanchez2025Futures}. By repeatedly positioning themselves as guardians of the future, corporate actors crowd out alternative imaginaries, reinforcing the notion that those who build these systems are best positioned to govern them.

HCI scholars must actively challenge these dynamics by rigorously scrutinizing corporate discourse, exposing power asymmetries, and shaping policy debates. Our analyses demonstrate that ostensibly participatory rhetoric often masks symbolic---rather than substantive---engagement, and HCI research is uniquely positioned to make such performative practices visible. Beyond analysis, research communities and conferences have a responsibility to critically examine their own institutional practices, such as sponsorship policies, to safeguard independence and uphold the integrity of participatory AI governance research, following precedents like the FAccT conference’s restrictions on corporate sponsorship \cite{barocas2023fact}.

For the HCI community, this implies that designing more collaborative or participatory governance processes is insufficient if researchers do not also interrogate the discursive conditions under which participation is invited, structured, and constrained. Our findings suggest that HCI research on AI governance should more explicitly analyze how corporate rhetoric shapes regulatory imagination, stakeholder roles, and the perceived limits of intervention. By treating discourse itself as a site of governance, HCI scholars can better identify where participatory ideals risk being co-opted—and where alternative sociotechnical arrangements might meaningfully redistribute power.

\section{Limitations and Future Work}

Our study has several limitations that future work can bridge and extend upon. First, our corpus consists exclusively of publicly available documents and pages produced by the three companies. We did not include internal documentation; as a result, our analysis captures primarily how companies communicate about AI safety externally, rather than the full range of internal practices or informal discussions. While this focus allows us to examine corporate narratives and public-facing discourse, it may not reflect all mechanisms through which safety practices are operationalized. Future work could combine document analysis with interviews, ethnography, or “read-along” studies to better understand how users interpret, navigate, and make sense of corporate safety narratives, as well as include ethnographic work with developers inside these companies. Additionally, the documents were often all-encompassing, speaking interchangeably about both models and platforms. While we retained this in our analysis, future research could more clearly differentiate between them to examine whether particular ideas are communicated through specific artifacts.

Second, we sampled three major platforms, excluding smaller companies, domain-specific systems, and materials beyond our keywords. Broader datasets could reveal additional discursive patterns and framings. 

Third, critical discourse analysis is interpretive; despite iterative coding and collaborative checks, our readings remain situated and cannot be generalizable. Complementary methods such as quantitative text analysis, surveys, or controlled experiments could assess how audiences perceive and act on safety discourse. Finally, corporate safety discourse evolves rapidly; longitudinal studies are needed to track changes alongside regulation, market pressures, and public debates.

\section{Conclusion}

This study offers one of the first empirical examinations of how safety is discursively constructed by corporate actors of generative AI platforms. Using CDA, we analyzed a stratified set of public-facing documents from OpenAI, Google, and Anthropic, focusing on how safety and other similar constructs were discursively constructed. By situating these materials in real-world contexts where users seek information about AI safety, our approach illuminated not only what companies claim about safety, but also how these claims are structured, framed, and legitimized through language.

Our analysis of corporate AI safety discourse reveals how companies construct authority, diffuse accountability, and frame risks, often positioning themselves as the primary stewards of safety while rhetorically invoking collaboration and participation. These findings underscore two key contributions for HCI: first, advancing AI literacy by treating corporate narratives as pedagogical artifacts that should be critically interrogated to understand assumptions, power dynamics, and ethical implications; and second, informing AI governance by showing how participatory language and metaphors can legitimize authority while constraining meaningful oversight or participation.


\begin{acks}
We thank Dr. Kelley Cotter, Dr. Elissa Redmiles, and Dr. Abraham Mhaidli for their invaluable feedback on earlier drafts of this work. We also appreciate the insights and support from our colleagues at the Max Planck Institute for Security and Privacy (MPI-SP), particularly members of the Human-Centered Security and Privacy group, and the LIKED Lab at Penn State. This research was partially funded by the Deutsche Forschungsgemeinschaft (DFG, German Research Foundation) under Germany's Excellence Strategy -- EXC 2092 CASA -- 390781972.
\end{acks}
\bibliographystyle{ACM-Reference-Format}
\bibliography{references.bib}

\appendix
\clearpage
\section{Appendix: Texts Cited from Corpus}

\renewcommand{\arraystretch}{1.2} 
\begin{table}[h!]
\centering
\begin{tabular}{|c|p{10cm}|}
\hline
\textbf{ID} & \textbf{Link} \\
\hline
15  & \url{https://openai.com/index/our-approach-to-ai-safety/} \\
\hline
29  & \url{https://www.anthropic.com/news/our-approach-to-understanding-and-addressing-ai-harms} \\
\hline
4   & \url{https://openai.com/business-data/} \\
\hline
19  & \url{https://openai.com/index/openai-safety-update/} \\
\hline
43  & \url{https://www.anthropic.com/news/reflections-on-our-responsible-scaling-policy} \\
\hline
45  & \url{https://www.anthropic.com/research/sabotage-evaluations} \\
\hline
3   & \url{https://openai.com/index/openai-board-forms-safety-and-security-committee/} \\
\hline
63  & \url{https://ai.google/static/documents/perspectives-on-issues-in-ai-governance.pdf} \\
\hline
68  & \url{https://policies.google.com/terms/generative-ai/use-policy} \\
\hline
47  & \url{https://www.anthropic.com/news/the-case-for-targeted-regulation} \\
\hline
35  & \url{https://www.anthropic.com/news/core-views-on-ai-safety} \\
\hline
58  & \url{https://blog.google/technology/ai/why-we-focus-on-ai-and-to-what-end/} \\
\hline
44  & \url{https://www.anthropic.com/news/anthropics-responsible-scaling-policy} \\
\hline
116 & \url{https://deepmind.google/discover/blog/taking-a-responsible-path-to-agi/} \\
\hline
34  & \url{https://www.anthropic.com/news/constitutional-classifiers} \\
\hline
33  & \url{https://www.anthropic.com/research/clio} \\
\hline
121 & \url{https://openai.com/charter/} \\
\hline
40  & \url{https://www.anthropic.com/news/national-security-expert-richard-fontaine-appointed-to-anthropic-s-long-term-benefit-trust} \\
\hline
27  & \url{https://www.anthropic.com/news/a-new-initiative-for-developing-third-party-model-evaluations} \\
\hline
36  & \url{https://www.anthropic.com} \\
\hline
9 & \url{https://openai.com/safety/how-we-think-about-safety-alignment/}\\
\hline
21 & \url{https://openai.com/transparency-and-content-moderation/}\\
\hline
126 & \url{https://blog.google/outreach-initiatives/public-policy/google-us-ai-action-plan-comments/}\\
\hline
121& \url{https://openai.com/charter/}\\
\hline
\end{tabular}
\end{table}

\end{document}